\documentstyle[12pt]{article}
\def\ts{\thinspace}

\newcommand\beq{\begin{equation}}
\newcommand\eeq{\end{equation}}
\newcommand{\rref}[1]{(\ref{#1})}
\newcommand\bear{\begin{array}}
\newcommand\enar{\end{array}}
\newcommand\Bear{\begin{eqnarray}}
\newcommand\Enar{\end{eqnarray}}
\newcommand\Bears{\begin{eqnarray*}}
\newcommand\Enars{\end{eqnarray*}}

\def\CMP{{\it Comm.\ts Math.\ts Phys.\ts}}

\def\FAP{{\it Funct.\ts Analy.\ts Appl.\ts}}
\def\IJMP{{\it Int.\ts J.\ts Mod.\ts Phys.\ts}}

\def\JP{{\it J.\ts Phys.\ts}}

\def\NP{{\it Nucl.\ts Phys.\ts}}
\def\PL{{\it Phys.\ts Lett.\ts}}

\def\TMP{{\it Theor.\ts Math.\ts Phys.\ts}}

\def\Zm{Zamolodchikov}
\def\AZm{A.\ts B.\ts \Zm}
\def\dur{H.\ts W.\ts Braden, E.\ts Corrigan, P.\ts E.\ts Dorey and R.\ts
Sasaki}
\begin{document}
\begin{flushleft}
{\it Yukawa Institute Kyoto}
\begin{flushright}
\begin{tabular}{l}
YITP/K-1091\\  YITP/U 95-08 \\
March 1995
\end{tabular}
\end{flushright}
\end{flushleft}

\begin{center}
\vspace*{1.0cm}

{\LARGE{\bf  Boundary Effects in Integrable Field Theory on a Half Line}}

\vskip 1.5cm

{\large {\bf Akira FUJII}}

\vskip 0.5 cm

{\sl Yukawa Institute for Theoretical Physics,} \\
{\sl Kyoto University,} \\
{\sl Kyoto 606-01, Japan.}
\vskip 0.5 cm
and

\vskip 0.5 cm

{\large {\bf Ryu SASAKI
\footnote{Supported partially by the grant-in-aid for Scientific
Research,
Priority Area 231 ``Infinite Analysis'' and General Research (C) in Physics,
Japan Ministry of Education.}}}

\vskip 0.5cm

{\sl Uji Research Center, Yukawa Institute for Theoretical Physics,} \\
{\sl Kyoto University,} \\
{\sl Uji 611, Japan.}

\end{center}

\vspace{1 cm}

\begin{abstract}
Boundary effects caused by the boundary interactions in various integrable
field theories on a half line are discussed at the classical as well as the
quantum level.
Only the so-called ``integrable" boundary interactions are discussed.
They are obtained by the requirement that certain combinations of the lower
members of the infinite set of conserved quantities should be preserved.
Contrary to the naive expectations, some ``integrable" boundary interactions
can drastically change the character of the theory.
In some cases,  for example, the sinh-Gordon theory,
 the theory becomes ill-defined because of the instability introduced by
 ``integrable" boundary interactions for a certain range of the parameter.
\end{abstract}

\vspace{1 cm}

\baselineskip=12pt 
\section{Introduction}
\setcounter{equation}{0}

Integrable field theory in 1 dimension or on a whole line,
both at the classical
and quantum levels, have been investigated quite intensively
in recent years and
many interesting results have been uncovered. The motivation is mainly
two-fold:
firstly, the integrable theory in its own right as a theoretical laboratory to
study the structure of field theory beyond perturbation and secondly its
connection with conformal field theory (deformed CFT) and string theory.

In contrast, integrable field theory on a half line, say $x\leq 0$,
 has a shorter history.
Here, the effects of the boundary or the boundary conditions which replace
the ``asymptotic conditions" in field theory on a whole line are the main
objects of research.
This problem is also related with integrable statistical lattice
models with non-trivial boundary conditions, scattering of
electrons by an impurity in solids (the Kondo problem) monopole catalysed
baryon decays and
deformations of conformal field theory with boundaries.
One of the purposes of this paper is to discuss the effects caused by
the boundary on the integrable systems.
Contrary to the naive expectations, some ``integrable" boundary interactions
can drastically change the character of the theory.
In some cases,  for example, the sinh-Gordon theory,
 the theory becomes ill-defined because of the instability introduced by a
specific choice of ``integrable" boundary interactions.
This will be discussed
in some detail in later sections.

There are two types of approaches for the integrable field theory on a half
line,
 algebraic and field theoretical. The algebraic approach was initiated by
Cherednik \cite{Cher}\ts more than ten years ago. Firstly, the dynamical system
under
consideration is integrable on the  whole line having a factorisable S-matrix.
Secondly, a natural  assumption is that when restricted to the
half line, {\em the particle content (mass spectrum), and the S-matrices
describing their mutual interactions, are exactly the same as those on
the whole line}.
Thirdly, when a particle hits the boundary it is assumed to be reflected
elastically (up to  rearrangements among  mass degenerate particles).
The compatibility of the reflections and the scatterings constitutes the
main algebraic condition, called Reflection equation \cite{Cher,GZa,Skl,GMN}\ts
which  generalises
 the Yang-Baxter
equation. In other words, the effect of the boundary is local and coded into
a set of reflection factors $K_{ab}(\theta )$,
where $a$ labels the incoming particle, $b$ labels the reflected particle and
 $\theta$ is the rapidity.

In the case of integrable field theory with {\em diagonal} S-matrices,
for example, affine Toda field theory to be discussed in detail in this paper,
the Yang-Baxter equation is trivially satisfied.
 The reflection at the boundary does not change the particle species and the
Reflection equation is again trivially satisfied. In this case a new algebraic
equation called the Bootstrap equation governs the exact S-matrices and the
Reflection equation is replaced by Reflection Bootstrap equation \cite{FK,RSb},
which
constrains  elastic reflection factors $K_a(\theta )$.

While the algebraic approach is applicable to any theories with
exact S-matrices, the field theoretical approach is useful for theories having
Lagrangians and classical equations of motion.
Our main concern in this paper is to discuss the boundary effects in the
field theoretic approach.
 Here the guiding principle is the
infinite set of conserved quantities which guarantees the integrability on a
whole line.
The boundary potential or the boundary interaction is so chosen as
to preserve the set of conserved quantities or  its suitable subset.
 Then the
boundary effects can be deduced from the explicit forms of the boundary
interactions  and field theoretical methods at the classical
and/or the quantum levels.
For well known integrable field theories on a whole line, sine-Gordon,
non-linear Schr\"odinger and affine Toda field theories, the ``integrable"
boundary interactions are deduced and/or conjectured \cite{Skl,Tar,CDRS,CDR}.
 It should be emphasised,
however, that not all of these ``integrable" boundary interactions give an
integrable field theory on a half line. Namely, preserving some infinite subset
of conserved quantities is a necessary but not  sufficient condition.
Compatibility with the other principles of field theory, in particular, at the
quantum level, must be checked
carefully.

We will discuss the boundary effects in integrable field theory on
a half line mostly taking explicit examples from affine Toda field theories.
This is because affine Toda field theory is one of the best understood
integrable field theory both at the classical and quantum levels.
The algebraic as well as the field theoretical approaches to affine Toda field
theory on a whole line have been very successful.
This paper is organised as follows:
in section 2 affine Toda field theory  is briefly reviewed in order to set the
stage and to introduce notation.  In section 3 very simple examples of a
harmonic oscillator with a negative  spring constant and a 1 dimensional wave
(string) with ``integrable" boundary interactions are discussed. It is shown
that the instability caused by a special ``integrable" boundary interaction
makes the theory ill-defined.
 In section 4 the boundary effects in the simplest
and best known affine Toda field theories, the sinh-Gordon and sine-Gordon
theories are discussed. Here it is shown that for a certain range of the
parameter in the ``integrable" boundary interaction the theory possesses
instability. In section 5 the boundary effects in affine Toda field theory,
mainly the $a_n$ series are discussed. Section 6 is for summary and discussion.


\section{Affine Toda field theory on a half line}
\setcounter{equation}{0}

Affine Toda field
theory \cite{MOPa} is a
massive scalar field theory with exponential interactions in $1+1$
dimensions
described by the action
\begin{equation}
S=\int dt\int_{-\infty}^\infty\ts dx{\cal L}.\label{wholeaction}
\end{equation}
Here the Lagrangian density is given by
\begin{equation}
{\cal L}={1\over 2}
\partial_\mu\phi^a\partial^\mu\phi^a-V(\phi )\label{ltoda}
\end{equation}
in which
\begin{equation}
V(\phi )={m^2\over
\beta^2}\sum_0^rn_ie^{\beta\alpha_i\cdot\phi}.\label{vtoda}
\end{equation}
Here $\phi$ is an $r$-component scalar field, $r$ is the rank of a
compact semi-simple Lie algebra $g$ with $\alpha_i$;
$i=1,\ldots,r$ being its simple roots. The roots are normalised so that long
roots have length 2, $\alpha_L^2=2$. An additional root,
$\alpha_0=-\sum_1^rn_i\alpha_i$ is an integer linear combination of the simple
roots, is called the affine root;
 it corresponds to the extra spot on an extended Dynkin-Kac diagram
for $\hat g$ and $n_0=1$.
When the term containing the extra root is removed, the theory becomes
conformally invariant (conformal Toda field theory).
The simplest affine Toda field theory, based on the simplest Lie algebra
$a_1$, the algebra of $su(2)$, is called sinh-Gordon theory, a cousin
of
the well known sine-Gordon theory. $m$ is a real parameter setting
the mass scale of the theory and $\beta$ is a real coupling constant,
which is relevant only in quantum theory.

Toda field theory is integrable at the classical level due to
the presence of an infinite number of conserved quantities.
Many beautiful properties of Toda field theory, both at the classical
and quantum levels, have been uncovered in recent years.
In particular, it is firmly believed that the integrability survives
quantisation.
The exact quantum S-matrices are known
\cite{AFZa,BCDSa,BCDSc,CMa,DDa}\cite{DGZc,CDS}
 for all the
Toda field theories based on non-simply laced algebras as well as those
based on simply laced algebras.
The singularity structure of the latter S-matrices, which in some
cases contain poles up to 12-th order \cite{BCDSc},
is  beautifully explained in terms of the
singularities of the corresponding Feynman diagrams
\cite{BCDSe}, so called Landau singularities.

 For the theory on a half line, \rref{wholeaction} will be replaced by
\begin{equation}
S=\int dt\left[\int_{-\infty}^0{\cal L}dx -{\cal
B}\right],\label{halfaction}
\end{equation}
 where ${\cal B}$, a function of the fields but not
their derivatives, represents the boundary interaction.
The stationarity condition of the action implies the equation of motion
which is the same as on the whole line and the boundary condition at
$x=0$
\begin{equation}
{\partial\phi\over\partial x}=-{\partial{\cal B}
\over \partial\phi}.\label{todaboundary}
\end{equation}
For any choice of ${\cal B}$, the energy
\begin{equation}
{\cal
E}=\int_{-\infty}^0\left[{1\over2}(\partial_t\phi^a)^2+{1\over2}(\partial_x\phi^a)^2+V(\phi)\right]dx+{\cal B}\label{Toda energy}
\end{equation}
is always conserved. But it is {\em no longer positive definite} for negative
boundary interaction (${\cal B}<0$).

In \cite{CDRS,CDR}\ts it was conjectured based on the analysis of low spin
conserved quantities that the generic form of the
``integrable" boundary interaction is given by
\begin{equation}
{\cal B}={m\over \beta^2}\sum_0^rA_ie^{{\beta\over 2}
\alpha_i\cdot\phi},\label{allboundary}
\end{equation}
where the coefficients $A_i,\ i=0,\dots ,r$ are a set of real numbers.
The condition
\rref{allboundary} is  a generalisation of the well known results
for the sine-Gordon theory ($r=1$), in which case the coefficients
$A_1$ and $A_0$ are completely arbitrary. However, for the affine Toda
field theory for higher rank algebras ($r\ge2$),
the coefficients are severely constrained due to the presence of higher spin
conserved quantities. For example, for the affine Toda field theories based
upon
the
$a_n^{(1)}$ series of Lie algebras the sequence of conserved charges includes
all spins (except zero) modulo $n+1$.
 Except for $a_1^{(1)}$, which corresponds
to the sinh-Gordon theory, each of these theories has conserved
charges of spin $\pm 2$. It was shown in \cite{CDRS} that
a combination of spin $\pm 2$  conserved
charges as well as  spin $\pm 3$  conserved
charges are preserved in the
presence of the
boundary interaction if the boundary interaction term has the form
\rref{allboundary} with  the further constraint:
\begin{equation}
\hbox{\bf either}\ A_i=\pm2, \ \hbox{for}\ i=0,\dots ,n\
\hbox{\bf or}\ A_i=0\ \hbox{for}\ i=0,\dots ,n\ .\label{anboundary}
\end{equation}
By similar analysis, a more general conjecture is obtained which applies to
all simply-laced affine Toda field theories
\cite{CDR,BCDR}\ts
\begin{equation}
\hbox{\bf either}\ A_i=\pm2\sqrt{n_i}, \ \hbox{for}\ i=0,\dots ,r\
\hbox{\bf or}\ A_i=0\ \hbox{for}\ i=0,\dots ,r\ .\label{simplacedboundary}
\end{equation}

The sine- and sinh-Gordon theories and possibly the Bullough-Dodd theory
(based on the $a_2^{(2)}$ algebra)  and other non-simply laced theories
\cite{BCDR,AF}
seem to be
the only ones for which there is a continuum of possible ``integrable" boundary
interactions. We will show in section 4 that some part of the continuum might
not be realised in field theory. For the others, the possible ``integrable"
boundary interactions consist of a choice of signs.


\section{Simple examples of instability caused by boundary}
\setcounter{equation}{0}

Let us start with a very simple example of a dynamical system with one degree
of freedom, namely a harmonic oscillator with an arbitrary spring constant $k$,
which is either {\em positive} or {\em negative}
\begin{equation}
L={1\over2}({dy(t)\over dt})^2 -{1\over2}k(y(t))^2.\label{harmosci}
\end{equation}
In either case, the system has one conserved quantity, the energy
\begin{equation}
{\cal E}={1\over2}({dy(t)\over dt})^2 +{1\over2}k(y(t))^2,\label{harmener}
\end{equation}
and satisfies the formal criterion of ``integrability". The solutions of the
equation of motion
\begin{equation}
{d^2y\over dt^2}=-ky
\end{equation}
are oscillatory for {\em positive} spring constant but they grow or decrease
exponentially
for {\em negative} spring constant $k$
\begin{equation}
y(t)=\pm e^{\pm \omega (t-t_0)}, \quad \omega^2=-k,\quad \hbox{for}\ \
k<0;\quad t_0\ \ \hbox{arbitrary}.\label{expsol}
\end{equation}
Namely the system is unstable for negative $k$ and hardly qualifies to be
called integrable in spite of the existence of the conserved energy.
In this case the energy is {\em no longer positive definite} and fails to
constrain
the system. In fact it is easy to see that these unstable exponential
solutions have {\em zero energy}.
The existence of non-trivial zero energy solutions (vacuum
solutions) would imply that the system can undergo certain changes without
costing energy, almost synonymous to instability.
A {\em negative energy solution} is also easily obtained
\begin{equation}
y(t)=e^{x_0}\cosh\omega t, \quad x_0\ \ \hbox{arbitrary}.\label{negosci}
\end{equation}
The energy can be made as large and negative as desired by choosing
large $x_0$.
 A quantum version of such
systems, if any, would have serious difficulties.

Next let us discuss the simplest integrable field theory on a half line, namely
a massless field (string) with a quadratic boundary potential,
\begin{equation}
{\cal L}={1\over2}\left[({\partial\phi(x,t)\over{\partial
t}})^2-({\partial\phi(x,t)\over{\partial x}})^2\right],\quad {\cal
B}={1\over2}A\phi^2(0,t).\label{quadbound}
\end{equation}
The equation of motion is the ordinary wave equation
\begin{equation}
\partial^2_t\phi-\partial^2_x\phi=0\label{waveeq}
\end{equation}
and the boundary condition is
\begin{equation}
{\partial\phi(x,t)\over{\partial x}}\Bigm|_{x=0}=-A\phi(0,t).\label{linbound}
\end{equation}
This is a linear system,
with a conserved energy
\begin{equation}
{\cal E}=
\int_{-\infty}^0\left[{1\over2}(\partial_t\phi)^2+{1\over2}(\partial_x\phi)^2\right]dx
+{1\over2}A\phi^2(0,t).\label{waveenergy}
\end{equation}
Therefore it is ``integrable" in the same sense as the harmonic oscillator.

The meaning of the boundary
interaction is now clear. It attaches a spring with spring constant $A$, which
is positive or negative, to the endpoint of the string. If $A<0$ the energy is
{\em no
longer positive definite} and unstable solutions exist:
\begin{equation}
\phi(x,t)=\pm e^{-A(x\pm t-x_0)},\quad \hbox{for}\ \ A<0.\label{waveunstab}
\end{equation}
They are finite everywhere on the half line $x\leq0$ for finite time $t$ and
localised near the boundary. It is again very easy to see that they have {\em
zero energy}. A {\em negative energy solution} is easily obtained
\begin{equation}
\phi(x,t)=e^{-A(x-x_0)}\cosh At , \quad x_0\
\hbox{arbitrary}.\label{negstring}
\end{equation}

It is easy to find a `plane wave' basis satisfying the boundary condition
\begin{equation}
u_p(x)\propto (ip -A)e^{ipx}+(ip+A)e^{-ipx}.\label{planebasis}
\end{equation}
It is also elementary to show that the above `plane wave' basis is orthogonal
to the localised solution $e^{-Ax}$ for $A<0$. This means that the `plane wave'
basis is not complete for $A<0$. Therefore the initial value problem
\begin{equation}
\phi(x,t=0)=F(x),\quad \partial_t\phi(x,t=0)=G(x), \quad x\leq0\label{iniprob}
\end{equation}
for the
string on the half line with the boundary is {\em unstable} for $A<0$ unless
$F(x)$ and $G(x)$ are exactly orthogonal to
the localised solution $e^{-Ax}$. A quantum version of such a theory, if any,
would meet serious difficulties. This simple example shows clearly that some
boundary effects, even if they are ``integrable", are not ``local" and can make
the theory ill-defined by the instability.

One can easily get the reflection factor \cite{CDRS,GMN}\ of the boundary from
the `plane wave' basis
\begin{equation}
K(p)={ip+A\over{ip-A}}\label{simpleref}
\end{equation}
for either sign of $A$. But for $A<0$ such a result seems superficial because
of the instability of the theory.

It is easy to note that adding a mass to the field tends to stabilise the
theory, as the mass term simply gives an attractive harmonic potential with a
spring constant $m^2$ at each point:
\begin{equation}
{\cal
L}={1\over2}\left[(\partial_t\phi(x,t))^2-(\partial_x\phi(x,t))^2-m^2\phi(x,t)^2\right],\quad
{\cal B}={1\over2}A\phi^2(0,t).\label{KGbound}
\end{equation}
Therefore the {\em zero energy} unstable mode exists only for $A\leq-m$
\begin{equation}
\phi(x,t)=e^{\pm\omega t-A(x-x_0)},\quad \omega^2=A^2-m^2.\label{KGzero}
\end{equation}
The latter condition is a disguise of the mass shell condition. A {\em negative
energy solution} is given by
\begin{equation}
\phi(x,t)=\cosh\omega t\ts e^{-A(x-x_0)},
\end{equation}
and the energy is a function of $x_0$, ${\cal E}={\omega^2\over{4A}}e^{2Ax_0}$,
which can be as large and negative as $x_0\to-\infty$. The `plane wave' basis
and
the reflection factor have the same form as the wave case.


\section{Sinh- and Sine-Gordon theories}
\setcounter{equation}{0}

Next let us consider sinh-Gordon theory, the simplest member of the  affine
Toda
field theories, based on the $a_1^{(1)}$ algebra. The Lagrangian density and
the boundary interactions in the notation of \rref{ltoda}\ts are
\begin{equation}
{\cal
L}={1\over2}\left[(\partial_t\phi)^2-(\partial_x\phi)^2\right]-{m^2\over{\beta^2}}
\left(e^{\sqrt2\beta\phi}+e^{-\sqrt2\beta\phi}\right),\ {\cal
B}={mA\over{\beta^2}}
\left(e^{\beta\phi/\sqrt2}+e^{-\beta\phi/\sqrt2}\right).\label{sinhlag}
\end{equation}
Here the parameter $A$ is arbitrary. Since the exponential functions are always
positive, it is expected that the boundary interaction term would induce strong
instability for large and negative $A$. We show this by constructing explicit
classical solutions for the equation of motion
\begin{equation}
\partial_t^2\phi-\partial_x^2\phi=-2\sqrt2{m^2\over\beta}\sinh\sqrt2\beta\phi,
\label{sinh-Gordoneq}
\end{equation}
with the boundary condition
\begin{equation}
\partial_x\phi\Bigm|_{x=0}=-\sqrt2{mA\over\beta}\sinh\beta\phi/\sqrt2\Bigm|_{x=0}.
\label{sinhbc}
\end{equation}
It is elementary to see that for $A<0$
\begin{equation}
\tanh{\sqrt2\beta\phi(x,t)\over4}=e^{\pm\omega t}e^{-mA(x-x_0)},\quad x_0>0
\label{sinhsol}
\end{equation}
are solutions provided
\begin{equation}
\omega^2=m^2(A^2-4)\geq0.\label{sinhonshell}
\end{equation}
For the positive sign, the r.h.s. of \rref{sinhsol}\ts eventually exceeds 1,
which is not allowed for tanh function with real arguments.

Namely, for $A\leq-2$ they are real and unstable solutions with one arbitrary
parameter $x_0>0$. It is also elementary to check that they are {\em zero
energy solutions}. For this the conserved energy takes the form
\begin{eqnarray}
{\cal
E}&=&\int_{-\infty}^0\left[{1\over2}(\partial_t\phi)^2+
{1\over2}(\partial_x\phi)^2+{m^2\over{\beta^2}}
\left(e^{\sqrt2\beta\phi}+e^{-\sqrt2\beta\phi}-2\right)\right]dx\nonumber\\
& &+{mA\over{\beta^2}}
\left(e^{\beta\phi/\sqrt2}+e^{-\beta\phi/\sqrt2}-2\right)\Bigm|_{x=0},
\label{sinhenergy}
\end{eqnarray}
in which constants are adjusted so that the trivial ``vacuum solution" $\phi=0$
has zero energy. This analysis also clarifies the dynamical meaning of the
$A=-2$
condition \rref{anboundary}\ts, which is the critical point for the
instability.
It should be remarked that for $A=-2$ there exists a one-parameter ($x_0$)
family of {\em time independent zero energy solutions} which could be
interpreted
as degenerate vacuua.
There are also {\em negative energy solutions} for $A<-2$ . They are obtained
by solving an initial value problem ($x_0>0$)
\begin{equation}
\phi(x,t=0)={4\over{\sqrt2\beta}}\hbox{arctanh} e^{-mA(x-x_0)},\quad
\partial_t\phi(x,t=0)=0.\label{negsinh}
\end{equation}
It is easy to calculate the energy
\begin{equation}
{\cal E}={2m\over{\beta^2}}(A-{4\over A}){a\over{1-a}},\quad
a=e^{2mAx_0},\label{shboundless}
\end{equation}
which goes to $-\infty$ as $x_0\to0$. In contrast, energy is bounded for
$A\geq-2$ \cite{CDR}.

  Since sinh- and sine-Gordon theories are
closely related, it is expected that sine-Gordon theory with ``integrable"
boundary interaction has instability for certain range of the parameter $A$.
 The Lagrangian density and
the boundary interactions for sine-Gordon theory are
\begin{equation}
{\cal
L}={1\over2}\left[(\partial_t\phi)^2-(\partial_x\phi)^2\right]
-{2m^2\over{\beta^2}}\left(1-\cos{\sqrt2\beta\phi}\right),\ \ {\cal
B}={2mA\over{\beta^2}}\left(1-\cos{\beta\phi/\sqrt2}\right).\label{sinelag}
\end{equation}
Here the parameter $A$ is arbitrary.  We have added constant terms to the
potential and boundary interaction term in
such a way that they vanish for the trivial ``vacuum solution" $\phi=0$. This
also
makes  them positive definite (up to the sign of $A$). Since the cosine  is a
bounded function, it is expected that the
instability caused by the boundary interaction term, if any, would be milder
than
the sinh-Gordon case. We show this by constructing explicit classical solutions
for the equation of motion
\begin{equation}
\partial_t^2\phi-\partial_x^2\phi=-2\sqrt2{m^2\over\beta}\sin\sqrt2\beta\phi,
\label{sine-Gordoneq}
\end{equation}
with the boundary condition
\begin{equation}
\partial_x\phi\Bigm|_{x=0}=-\sqrt2{mA\over\beta}\sin\beta\phi/\sqrt2\Bigm|_{x=0},
\label{sinebc}
\end{equation}
in the same way as in the sinh-Gordon theory.
It is elementary to see that
\begin{equation}
\tan{\sqrt2\beta\phi(x,t)\over4}=e^{\pm\omega t}e^{-mA(x-x_0)},\quad x_0\ \
\hbox{arbitrary}
\label{sinesol}
\end{equation}
are solutions provided
\begin{equation}
\omega^2=m^2(A^2-4)\geq0.\label{sineonshell}
\end{equation}

Namely, for $A^2\geq4$ they are ordinary kink solutions with a specific $x$
dependence. The one arbitrary
parameter $x_0$ is interpreted as the kink position. In sharp contrast with the
sinh-Gordon case, the field
$\phi$ is always finite for finite $x$ and $t$. However, for $A\leq-2$
(negative
boundary term)  they are {\em zero energy
solutions}, a sign of instability. For this the conserved energy takes the
form
\begin{eqnarray}
{\cal
E}&=&\int_{-\infty}^0\left[{1\over2}(\partial_t\phi)^2+
{1\over2}(\partial_x\phi)^2+{2m^2\over{\beta^2}}
\left(1-\cos{\sqrt2\beta\phi}\right)\right]dx\nonumber\\
& &+{2mA\over{\beta^2}}
\left(1-\cos{\beta\phi/\sqrt2}\right)\Bigm|_{x=0}\geq{4mA\over{\beta^2}},
\label{sineenergy}
\end{eqnarray}
in which constants are adjusted so that the trivial ``vacuum solution" $\phi=0$
has zero energy.
As in the case of sinh-Gordon theory, for $A=-2$ there exists a one-parameter
($x_0$) family of {\em time independent zero energy solutions} which could be
interpreted as degenerate vacuua. Negative energy solutions can also be
obtained by solving an initial value problem similar to \rref{negsinh}.
However, in contrast with the sinh-Gordon theory, the energy is bounded from
below.

Let us conclude this section by remarking that field theoretical formulation of
sinh- and/or sine-Gordon theories with ``integrable" boundary interaction for
$A\leq-2$ both at the classical and quantum levels seems problematic.
It might be possible to interpret the plethora of the solutions of Reflection
Bootstrap equation \cite{RSb}\ts in terms of the instability.
Further investigation is definitely wanted.


\section{$a_n^{(1)}$ Toda field theory}
\setcounter{equation}{0}

Among the $2^{n+1}$ possible choices of the ``integrable" boundaries for
$a_n^{(1)}$ Toda field theory \rref{anboundary}, here we discuss only two
cases,
namely all $A_i$ are the same
\begin{equation}
A_i=A=\pm 2,\quad i=0,1,\ldots,n.\label{Znsymbc}
\end{equation}
The other cases, with mixed pluses and minuses, break the $Z_{n+1}$ symmetry
which is essential for the determination of the bulk S-matrices
\cite{AFZa,BCDSc}. Moreover, in these cases, the ``vacuum" $\phi=0$ is no
longer
the solution of equation of motion and the boundary condition.
Therefore, the ``new vacuum", if any, of such theories is different from that
of
the whole line.
The particle spectrum and their interactions are also different due to the lack
of $Z_{n+1}$ symmetry and the different vacuum.

First we show that $a_{2n+1}^{(1)}$ theory with $A=-2$ ``integrable" boundary
interaction is {\em unstable} in spite of the $Z_{2n+2}$ symmetry.
We have shown this for $a_1^{(1)}$ theory, the sinh-Gordon case.
$a_{2n+1}^{(1)}$ theory with $A=-2$ has also a one parameter ($x_0$) family of
{\em zero energy solutions} (degenerate vacuua) of the same form as in the
sinh-Gordon theory:
\begin{eqnarray}
\phi(x,t)&=&\mu\varphi(x,t),\quad
\mu=\alpha_1+\alpha_3+\cdots+\alpha_{2n+1},\nonumber\\
\tanh{\beta\varphi(x,t)\over2}&=&e^{-A(x-x_0)}.\label{anredsol}
\end{eqnarray}
The single component field $\varphi$ in the special direction $\mu$ satisfies
all the $2n+1$ equations for $\phi$. One only has to note $\alpha_i\cdot\mu=2$
for $i$ odd and $-2$ for $i$
even. This phenomenon is called ``dimension one
reduction" and has been discussed in some detail in \cite{RSa}\ts for various
types of Toda field theory.
However, it should be remarked that the reductions applicable to the whole line
are not guaranteed to work for the boundary.
The existence of the zero (frequency) mode was also noted in \cite{CDRS}\ts
within the linear approximation. It should be remarked that the above solution
can also be obtained by using Hirota's method.

In contrast, $a_{\rm even}^{(1)}$ theory with $A=-2$ seems to have no
degenerate vacuua.
So it deserves further investigation. Let us consider the weak coupling limit
($\beta\ll1$), or the linear approximation.
Then the system is a sum of independent Klein-Gordon fields, with ``integrable"
quadratic boundary interactions:
\begin{eqnarray}
{\cal L}&=&\sum_{j=1}^{2n}|\partial_\mu\phi_j|^2-m_j^2|\phi_j|^2,\quad
m_j=2m\sin{j\pi\over{2n+1}},\nonumber\\
{\cal B}&=&-{1\over{2m}}\sum_{j=1}^{2n}m_j^2|\phi_j|^2.\label{linappr}
\end{eqnarray}
Due to the attractive boundary effect, each field has a localised solution,
$e^{{m_j^2\over{2m}}x}$, namely a boundary bound state.
As before this state is orthogonal to the `plane wave' basis
$$
u_p^{(j)}(x)\propto (ip
+{m_j^2\over{2m}})e^{ipx}+(ip-{m_j^2\over{2m}})e^{-ipx}.
$$
This means that upon quantisation, one needs to introduce the creation and
annihilation operators for the `plane wave' states as well as the boundary
bound states:
\begin{eqnarray}
\phi_j(x,t)&=&{1\over{2\pi}}\int_0^\infty{dp\over\sqrt{N_p}}\left\{e^{-i\omega_pt}a_j(p)+e^{i\omega_pt}a_{\bar
j}^\dagger(p)\right\}u_p^{(j)}(x) \nonumber\\
&+&{1\over\sqrt{N_j}}\left\{e^{-i\omega_jt}b_j+e^{i\omega_jt}b_{\bar
j}^\dagger\right\}e^{{m_j^2\over{2m}}x}.\label{quantise}
\end{eqnarray}
Here $N_p$ and $N_j$ are normalisation constants and $\omega_p^2=p^2+m_j^2$,
$\omega_j^2=m_j^2(1-{m_j^2\over{4m^2}})$.
Otherwise the Heisenberg commutation relations
$$
[\phi_j(x,t),\partial_t\phi_k(y,t)]=i\delta_{j\bar k}\delta(x-y)
$$
cannot be satisfied. Namely the spectrum is changed by the boundary.
Therefore a naive correspondence with the algebraic approach \`a la Cherednik
seems to be lost. However, it is tempting to relate the emergence of the
boundary bound states with the plethora of the solutions
of the Reflection Bootstrap equation.
Certain characteristic differences between the solutions of reflection
Bootstrap equation for $a_{\rm even}$ and $a_{\rm odd}$ theories are also
reported \cite{RSb}.

\section{Summary and discussion}
\setcounter{equation}{0}

The effects of the `boundary interactions' are analysed for various types of
integrable field theories on a half line.
Here we discuss only the ``integrable'' boundary interactions such that they
preserve certain subset of the classical infinite set of conserved quantities.
It is shown that not all of these `integrable' boundary interactions give
consistent quantum field theories on a half line.
Especially when the boundary interaction is negative and strong, some theories
become ill-defined due to the instabilities which are related with the
non-positive definite energy.

For sinh-Gordon theory with a strong negative boundary interaction,
an explicit 1-parameter ($x_0$) family of unstable solutions \rref{sinhsol}\ts
are constructed, which has the form of 1-soliton solution with
$x_0$  being its position.
It is well known that sinh-Gordon theory and other affine Toda field theories
on a whole line have  no soliton solutions. The negative boundary interaction
enables the soliton solution and thereby causes the instability.
It is also interesting that the unstable solution can be obtained from the
``trivial vacuum" $\phi=0$ by the B\"acklund Transformation (B-T).
The close similarity between the B-T and the ``integrable" boundary
interactions \cite{Tar,SSW} will be discussed elsewhere \cite{FSb}.

Although  we did not discuss the free boundary, ${\cal B}=0$, in this paper,
this does not mean the free boundary is uninteresting.
The free boundary  always preserves the necessary conserved quantities and
therefore is ``integrable" ,
not only for affine Toda field theory  but also for other types of theories
like non-linear sigma models which have non-diagonal S-matrices.
The corresponding Neumann boundary condition, $\partial_x\phi=0$ at $x=0$
is always satisfied if $\phi$ is extended to the whole line as an even
function.
It is an interesting challenge to derive reflection factors $K_a(\theta)$
corresponding to the free boundary for various models in terms of the field
theoretical methods.

\section*{Acknowledgments}
We thank Ed Corrigan and Q-H.\ts Park for fruitful discussion. R.\ts S is
grateful for Japan Society for Promotion of Sciences for financial support
which enabled him to visit UK.

\end{document}